\documentclass[aip, jcp, twocolumn,  reprint, citeautoscript, showpacs, amsmath, amssymb, superscriptaddress]{revtex4-1}

\usepackage{graphicx}
\usepackage{dcolumn}

\begin{document}

\title{Implicit self-consistent electrolyte model in plane-wave density-functional theory}

\author{Kiran Mathew}
\thanks{These two authors contributed equally.}\affiliation{Department of Materials Science and Engineering, Cornell University, Ithaca, New York 14853, USA}
\affiliation{Department of Materials Science and Engineering, University of Florida, Gainesville, Florida 32611, USA}

\author{V. S. Chaitanya Kolluru}
\thanks{These two authors contributed equally.}
\affiliation{Department of Materials Science and Engineering, University of Florida, Gainesville, Florida 32611, USA}
\affiliation{Quantum Theory Project, University of Florida, Gainesville, Florida 32611, USA}

\author{Srinidhi Mula}
\affiliation{Department of Materials Science and Engineering, University of Florida, Gainesville, Florida 32611, USA}
\affiliation{Quantum Theory Project, University of Florida, Gainesville, Florida 32611, USA}

\author{Stephan N. Steinmann}
\affiliation{Univ Lyon, Ecole Normale Sup\'erieure de Lyon, CNRS Universit\'e Lyon 1, Laboratoire de Chimie  UMR 5182, 46 all{\'e}e d'Italie,  F-69364, LYON, France}

\author{Richard G. Hennig} \email{rhennig@ufl.edu}
\affiliation{Department of Materials Science and Engineering, University of Florida, Gainesville, Florida 32611, USA}
\affiliation{Quantum Theory Project, University of Florida, Gainesville, Florida 32611, USA}


\begin{abstract}
  The ab-initio computational treatment of electrochemical systems requires an appropriate treatment of the solid/liquid interfaces. A  fully quantum mechanical treatment of the interface is computationally demanding due to the large number of degrees of freedom involved. In this work, we describe a computationally efficient model where the electrode part of the interface is described at the density-functional theory (DFT) level, and the electrolyte part is represented through an implicit solvation model based on the Poisson-Boltzmann equation.  We describe the implementation of the linearized Poisson-Boltzmann equation into the Vienna Ab-initio Simulation Package (VASP), a widely used DFT code, followed by validation and benchmarking of the method. To demonstrate the utility of the implicit electrolyte model, we apply it to study the surface energy of Cu crystal facets in an aqueous electrolyte as a function of applied electric potential. We show that the applied potential enables the control of the shape of nanocrystals from an octahedral to a truncated octahedral morphology with increasing potential.
\end{abstract}


\maketitle

\section{Introduction}
\label{sec:pbz_introduction}
Interfaces between dissimilar systems are ubiquitous in nature and frequently encountered and employed in scientific and engineering applications. Of particular interest are solid/liquid interfaces that form the crux of many technologically essential systems such as electrochemical interfaces \cite{abruna, robinson, npelectrochem}, corrosion, lubrication and nanoparticle synthesis.\cite{hanrath, npinterface, liu2012interconversion} A comprehensive understanding of the behavior and properties of such systems requires a detailed microscopic description of the solid/liquid interfaces.

Given the heterogeneous nature and the complexities of solid/liquid interfaces,\cite{agubra2014formation}  experimental studies often need to be supplemented by theoretical undertakings. A complete theoretical investigation of such interface based on an explicit description of the electrolyte and solute involves solving systems of nonlinear equations with a large number of degrees of freedom, which quickly becomes prohibitively expensive even when state of the art computational resources are employed.\cite{ta2} This calls for the development of efficient and accurate implicit computational frameworks for the study of solid/liquid interfaces.\cite{Galli2000, ta2, ta5, cramer, ta3, Marzari, Ringe2016}

\begin{figure}[t]
  \includegraphics[width=\columnwidth]{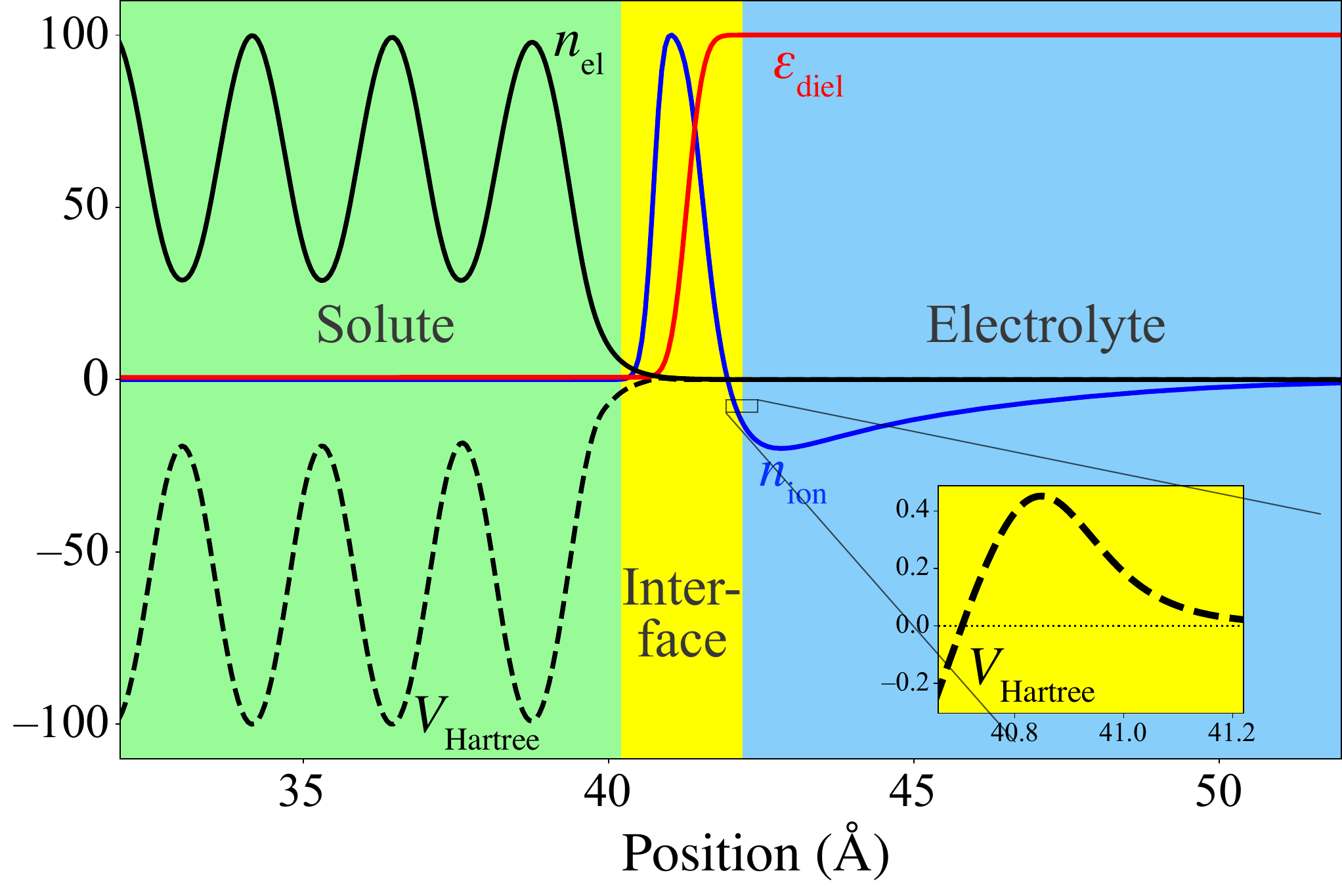}
  \caption{\label{fig:regions} Spatial decomposition of the solid/electrolyte system into the solute, interface, and electrolyte regions for a fcc Pt(111) surface embedded in an electrolyte with relative permittivity $\epsilon_r = 78.4$ and monovalent cations and anions with concentration 1 ~M. The solute is described by density-functional theory, the electrolyte by an implicit solvation model. The interface region is formed self-consistently as a functional of the electronic density of the solute. All properties are shown as a percentage of their maximum absolute value. The inset shows the relatively invisible peak in the Hartree potential at the interface.}
\end{figure}

We have previously developed a self-consistent implicit solvation model\cite{Mathew2014, VASPsol_GitHub} that describes the dielectric screening of a solute embedded in an implicit solvent, where the solute is described by density-functional theory (DFT). We implemented this solvation model into the widely used DFT code VASP.\cite{VASP} This software package, VASPsol, is freely available under the open-source Apache License, Version 2.0 and has been used by us and others to study the effect of the presence of solvent on a number of materials and processes, such as the catalysis of ketone hydrogenation,\cite{vsol_ketone} ionization of sodium superoxide in sodium batteries,\cite{vsol_jinsoo} the chemistry at electrochemical interfaces,\cite{vsol_lmno4,Steinmann2015a} hydrogen adsorption on platinum surfaces,\cite{vsol_gross} surface stability and phase diagram of Mica in solvent,\cite{vsol_vatti} {\it etc}.\cite{vsol_goodpaster, vsol_sakong}

In this work, we describe the extension of our solvation model VASPsol\cite{Mathew2014, VASPsol_GitHub} to include the effects of mobile ions in the electrolyte through the solution of the linearized Poisson-Boltzmann equation and then apply the method to electrochemical systems to demonstrate the power of this approach. Figure~\ref{fig:regions} illustrates how the solid/electrolyte system is divided into spatial regions that are described by the explicit DFT and the implicit solvent. This  solvation model allows for the description of charged systems and the study of the electrochemical interfacial systems under applied external voltage. Section~\ref{sec:pbz_framework} outlines the formalism and presents the derivation of the energy expression. Sec.~\ref{sec:pbz_implementation} describes the details of the implementation and Sec.~\ref{sec:pbz_validation} validates the method against previous computational studies of electrochemical systems. Finally, in Sec.~\ref{sec:pbz_applications} we apply the approach to study how an applied electric potential can control the equilibrium shape of Cu  nanocrystals.

\section{Theoretical framework}
\label{sec:pbz_framework}

Our solvation model couples an implicit description of the electrolyte to an explicit quantum-mechanical description of the solute. The implicit solvation model incorporates the dielectric screening due to the permittivity of the solvent and the electrostatic shielding due to the mobile ions in the electrolyte. The solute is described explicitly using DFT.

To obtain the energy of the combined solute/electrolyte system, we spatially divide the materials system into three regions: (i) the solute that we describe explicitly using DFT, (ii) the electrolyte that we describe using the linearized Poisson-Boltzmann equation and (iii) the interface region, that electrostatically couples the DFT and Poisson-Boltzmann equation. Fig.~\ref{fig:regions} illustrates the regions for a Pt(111) surface embedded in a 1~M aqueous electrolyte. The interface between the electrolyte and the solute system is formed self-consistently through the electronic density, and the interaction between the electrolyte and solute is described by the electrostatic potential as explained in the following sections.

\subsection{Total energy}

We define the interface region between the solute and the electrolyte using the electronic charge density of the solute, $n(\vec r)$. The electrolyte occupies the volume of the simulation cell where the solute's electronic charge density is essentially zero. In the interface region where the electronic charge density increases rapidly from zero towards the bulk values inside of the solute, the properties of the implicit solvation model, {\it e.g.}, the permittivity, $\epsilon$, and the Debye screening length, $\lambda_\mathrm{D}$, change from their bulk values of the electrolyte to the vacuum values in the explicit solute region. The scaling of the electrolyte properties is given by a shape function\cite{Mathew2014, ta4, ta5, ta9, ta11}
\begin{align}
  \zeta(n(\vec r)) =
  \frac{1}{2}\text{erfc}
  \left \{ \frac{\text{log}(n/n_\mathrm{c})}{\sigma \sqrt{2}} \right \}.
\end{align}

Following our previous work\cite{Mathew2014} and the work by Arias {\it et al.},\cite{ta4, ta5, ta9, ta11} the total free energy, $A$, of the system consisting of the solute and electrolyte is expressed as \begin{align}
  \label{eq:free-energy}
  A[n(\vec r), \phi(\vec r)] & = A_\mathrm{TXC}[n(\vec r)] \nonumber
  + \int\phi(\vec r) \rho_s (\vec r) \, d^3r \nonumber \\&
  - \int \epsilon(\vec r) \frac{| \nabla \phi |^2}{8 \pi} \, d^3r
  + \int \frac{1}{2}\phi (\vec r) \rho_\mathrm{ion} (\vec r) \, d^3r \nonumber \\
  &+ A_\mathrm{cav} + A_\mathrm{ion} ,
\end{align}
where $A_\mathrm{TXC}$ is the kinetic and exchange-correlation contribution from DFT, $\phi$ is the net electrostatic potential of the system, and $\rho_\mathrm{s}$ and $\rho_\mathrm{ion}$ are the total charge density of the solute and the ion charge density of the electrolyte, respectively.

The total solute charge density, $\rho_\mathrm{s}$, is the sum of the solute electronic and nuclear charge densities, $n(\vec r)$ and $N(\vec r)$, respectively,
\begin{equation}
  \rho_\mathrm{s}(\vec r) = n(\vec r) + N(\vec r).
\end{equation}
The ion charge density of the electrolyte, $\rho_\mathrm{ion}$, is given by
\begin{equation}
  \rho_\mathrm{ion}(\vec r) = \sum_i q z_i c_i (\vec r),
\end{equation}
where $c_i (\vec r)$ is the concentration of ionic species $i$,
$z_i$, denotes the charge state, and $q$ is the elementary charge. The
concentration of ionic species, $c_i$, is given by a Boltzmann factor
of the electrostatic energy\cite{honig} and modulated in the interface
region by the shape function, $\zeta(n(\vec r))$,
\begin{equation}
  c_i (\vec r) = \zeta[n(\vec r)] c_i^0 \exp\left(\frac{-z_i q \phi(\vec
    r)}{k_\mathrm{B} T} \right),
\end{equation}
where $c_i^0$ is the bulk concentration of ionic species $i$,
$k_\mathrm{B}$ the Boltzmann constant, and $T$ the temperature.

The relative permittivity of the electrolyte, $\epsilon(\vec r)$, is assumed to be a local functional of the electronic charge density of the solute and modulated by the shape function,\cite{Mathew2014, ta4,
  ta5, ta9, ta11}
\begin{equation}
  \epsilon(n(\vec r)) = 1 + (\epsilon_\mathrm{b} - 1) \zeta(n(\vec r)),
\end{equation}
where $\epsilon_\mathrm{b}$ is the bulk relative permittivity of the solvent.

The cavitation energy, $A_\mathrm{cav}$, describes the energy required to form the solute cavity inside the electrolyte and is given by
\begin{equation}
  \label{eq:tau}
  A_\mathrm{cav} = \tau \int |\nabla \zeta(n(\vec r))| d^3r,
\end{equation}
where the effective surface tension parameter $\tau$ describes the cavitation, dispersion, and the repulsion interaction between the solute and the solvent that are not captured by the electrostatic terms.\cite{Marzari}

In comparison to our previous solvation model,\cite{Mathew2014} the main changes to the free energy expression are the inclusion of the electrostatic interaction of the ion charge density in the electrolyte with the system's electrostatic potential, $\phi$, and the non-electrostatic contribution to the free energy from the mobile ions in the electrolyte, $A_\mathrm{ion}$. In our work we assume this non-electrostatic contribution from the ions to consist of just the entropy term,
\begin{equation}
  A_\mathrm{ion} = k_BTS_\mathrm{ion},
  \end{equation}
where $S_\mathrm{ion}$ is the entropy of mixing of the ions in the
electrolyte, which for small electrostatic potentials can be
approximated to first order as\cite{honig}
\begin{align}
  \label{eq:entropy}
  S_\mathrm{ion} &= \int \sum_i c_i \ln \left (\frac{c_i}{c_i^0}
  \right ) d^3r \nonumber \\ &\approx - \int \sum_i c_i \frac{z_i q
    \phi}{k_BT}d^3r.
\end{align}

\subsection{Minimization}

To obtain the stationary point of the free energy, $A$, given by Eq.~\eqref{eq:free-energy}, we set the first order variations of the free energy with respect to the system potential, $\phi$, and the electronic charge density, $n(r)$, to zero.\cite{Mathew2014, ta4, ta5, ta9, ta11} Taking the variation of $A[n(\vec r), \phi(\vec r)]$ with respect to the electronic charge density, $n(\vec r)$, yields the typical Kohn-Sham Hamiltonian\cite{ks} with the following additional term in the local part of the potential
\begin{align}
  \label{eq:local-potential}
  V_\mathrm{solv} =&
  \frac{\delta \epsilon(n)}{\delta n} \frac{| \nabla \phi |^2}{8 \pi} +
  \phi \frac{\delta \rho_\mathrm{ion}}{\delta n} \nonumber \\
  &+ \tau \frac{\delta |\nabla \zeta|}{\delta n} +
  k_BT \frac{\delta S_\mathrm{ion}}{\delta n}.
\end{align}
Taking the variation of $A[n(\vec r), \phi(\vec r)]$ with respect to
$\phi(\vec r)$ yields the generalized Poisson-Boltzmann
equation
\begin{equation}
  \vec{\nabla} \cdot \epsilon \vec{\nabla} \phi = - \rho_\mathrm{s}
  -\rho_\mathrm{ion},
\end{equation}
where $\epsilon(n(r))$ is the relative permittivity of the solvent as
a local functional of the electronic charge density.

We further simplify the system by considering the case of electrolytes
with only two types of ions present, whose charges are equal and
opposite, {\it i.e.} $c_1^0$ = $c_2^0$ = $c^0$ and $z_1$ = $-z_2$ =
$z$. Then the ion charge density of the electrolyte
becomes\cite{honig}
\begin{align}
  \label{eq:rho_ion}
  \rho_\mathrm{ion} & =  \zeta[n(\vec r)] q z c^0 \left [ \exp\left (\frac{-z q \phi}{k_BT} \right )
    - \exp \left (\frac{z q \phi}{k_BT} \right ) \right ] \nonumber \\
  &=  - 2 \zeta[n(\vec r)] q z c^0 \sinh\left (\frac{z q \phi}{k
  T} \right)
\end{align}
and the Poisson-Boltzmann equation becomes
\begin{equation}
  \label{eq:PBeq}
  \vec{\nabla} \cdot \epsilon \vec{\nabla} \phi =
  - \rho_\mathrm{s} +
  2 \zeta[n(\vec r)] q z c^0 \sinh \left (\frac{z q \phi}{k_BT} \right ).
\end{equation}
For small arguments $x = \frac{z q \phi}{k_BT} \ll 1$, $\sinh(x) \to x$
and we obtain the linearized Poisson-Boltzmann equation
\begin{equation}
  \label{eq:linPBeq}
  \vec{\nabla} \cdot \epsilon \vec{\nabla} \phi - \kappa^2 \phi = -
  \rho_\mathrm{s}
\end{equation}
with
\begin{equation}
  \label{eq:kappa}
  \kappa^2 = \zeta[n(\vec r)] \left(\frac{2 c^0 z^2 q^2}{k_BT} \right) =
  \zeta[n(\vec r)] \frac{1}{\lambda_\mathrm{D}^2},
\end{equation}
where $\lambda_\mathrm{D}$ is the Debye length that characterizes
the dimension of the electrochemical double layer.

The implicit electrolyte model is given by the additional local
potential, $V_\mathrm{solv}$, of Eq.~\eqref{eq:local-potential} and the
linearized Poisson-Boltzmann equation given by Eqs.~\eqref{eq:linPBeq}
and \eqref{eq:kappa}. This model has four key approximations: (i) The
ionic entropy term of Eq.~\eqref{eq:entropy} treats the ionic solution
as an ideal system and assumes that any interactions between the ions
are small compared to $k_BT$. (ii) The cations and anions of the
electrolyte have equal and opposite charge, simplifying the
Poisson-Boltzmann equation. (iii) The electrostatic potential in the electrolyte region is small, such
that $zq\Phi \ll k_BT$, which leads to the linear approximation of the
Poisson-Boltzmann equation~\eqref{eq:linPBeq}. To be quantitative, arguments to the sinh function in Eq.~\eqref{eq:PBeq} of less than 0.25 lead to errors below 1\%. Hence, in the region where the shape function is unity (and thus $\vec{\nabla} \cdot \epsilon \approx 0$), charge excesses of up to 0.5 $c^0$ are still faithfully reproduced. (iv) The ions
are treated as point charges. The finite size of the ions in solution
would limit their maximum concentration in the electrolyte, and the
model could lead to unphysically high concentrations near the
electrodes. These four approximations could be overcome and will be
considered in future extensions of the VASPsol model.

\section{Implementation}
\label{sec:implementation}

We implement the implicit electrolyte model described above into the widely used DFT software Vienna Ab-intio Software Package (VASP)~\cite{VASP}. VASP is a parallel plane-wave DFT code that supports both ultra-soft pseudopotentials~\cite{uspp,kresseuspp} as well as the projector-augmented wave (PAW)~\cite{paw} formulation of pseudopotentials. Our software module, VASPsol, is freely distributed as an open-source package and hosted on GitHub at \url{https://github.com/henniggroup/VASPsol}.\cite{VASPsol_GitHub}

The main modifications in the code are the evaluation of the additional contributions to the total energy and the local potential, given by Eqs.~\eqref{eq:free-energy} and \eqref{eq:local-potential}, respectively. Corrections to the local potential require the solution of the linearized Poisson-Boltzmann equation given by Eq.~\eqref{eq:PBeq} in each self-consistent iteration. Our implementation solves the equation in reciprocal space and makes efficient use of fast Fourier transformations (FFT).\cite{Mathew2014} This enables our implementation to be compatible with the Message Passing Interface (MPI) and to take advantage of the memory layout of VASP. We use a pre-conditioned conjugate gradient algorithm to solve the linearized Poisson-Boltzmann equation with the pre-conditioner $\left ( G^2 + \kappa^2 \right )^{-1}$, where $G$ is the magnitude of the reciprocal lattice vector and $\kappa^2$ is the inverse of the square of the Debye length, $\lambda_D$.

\begin{figure}[t]
  \includegraphics[width=\columnwidth]{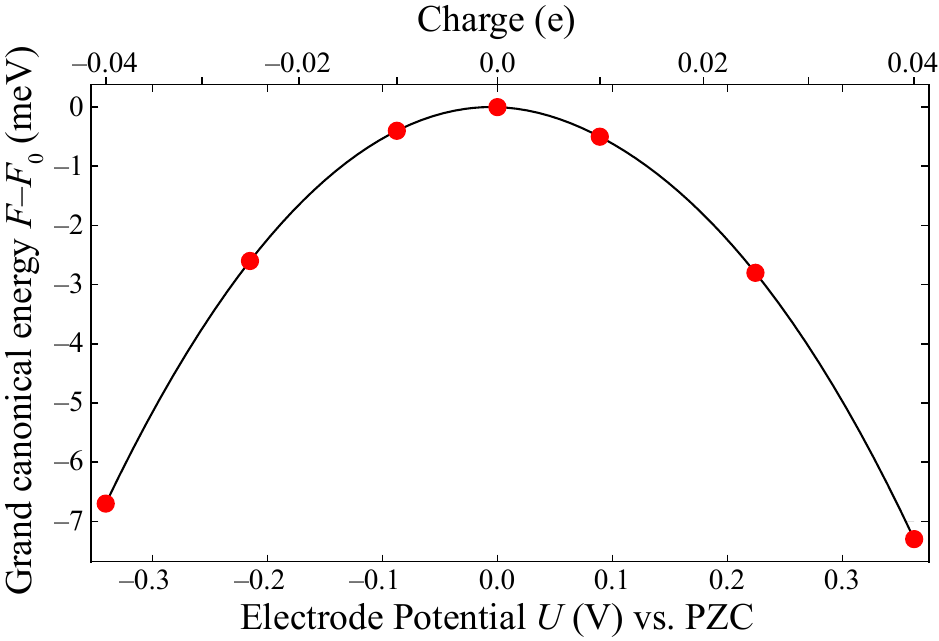}
  \caption{\label{fig:free-energy} The grand canonical electronic energy, $F(U)$ of a charged Pt(111) slab relative to the neutral slab as a function of the electrode potential $U$. The grand canonical energy displays the maximum correctly at the neutral slab when the reference electrostatic potential is chosen such that the potential is zero in the bulk of the electrolyte and the necessary correction due to the reference potential $\Delta E_\mathrm{ref} = n_\mathrm{electrode} \, \Delta U_\mathrm{ref}$ is included in the energy.}
\end{figure}

The solution of the linearized Poisson-Boltzmann equation~\eqref{eq:PBeq} provides a natural reference electrostatic potential by setting the potential to zero in the bulk of the electrolyte. However, plane-wave DFT codes, such as VASP, implicitly set the average electrostatic potential in the simulation cell to zero, not the potential in the electrolyte region. For simplicity, we do not modify the VASP reference potential but provide the shift in reference potential that needs to be added to the Kohn Sham eigenvalues and the Fermi level. Furthermore, the shift of the electrostatic potential to align the potential in the electrolyte region to zero, or to any other value, modifies the energy of the system. This energy change is given by $\Delta E_\mathrm{ref} = n_\mathrm{electrode} \, \Delta U_\mathrm{ref}$, where $n_\mathrm{electrode}$ is the net charge of the electrode slab and $\Delta U_\mathrm{ref}$ is the change in reference potential, {\it i.e.}, the shift to align the potential in the electrolyte region to zero.

To validate the change in reference potential, Fig.~\ref{fig:free-energy} shows the grand canonical electronic energy, $F(U)$, as a function of the applied potential, $U$, of the Pt (111) electrode slab.
The grand-canonical electronic energy, $F$, is the Legendre transformation of the free energy, A, of the system, $F(U) = A(n) - n_\mathrm{electrode}  U$, and is always lowered when charge is added or removed from the neutral slab. The grand canonical energy, $F$, exhibits the expected quadratic behavior~\cite{Santos2004} and the maximum at the neutral slab when the energy change $\Delta E_\mathrm{ref}$ due the change in reference potentials is included.

\label{sec:pbz_implementation}
\begin{figure}[t]
  \includegraphics[width=\columnwidth]{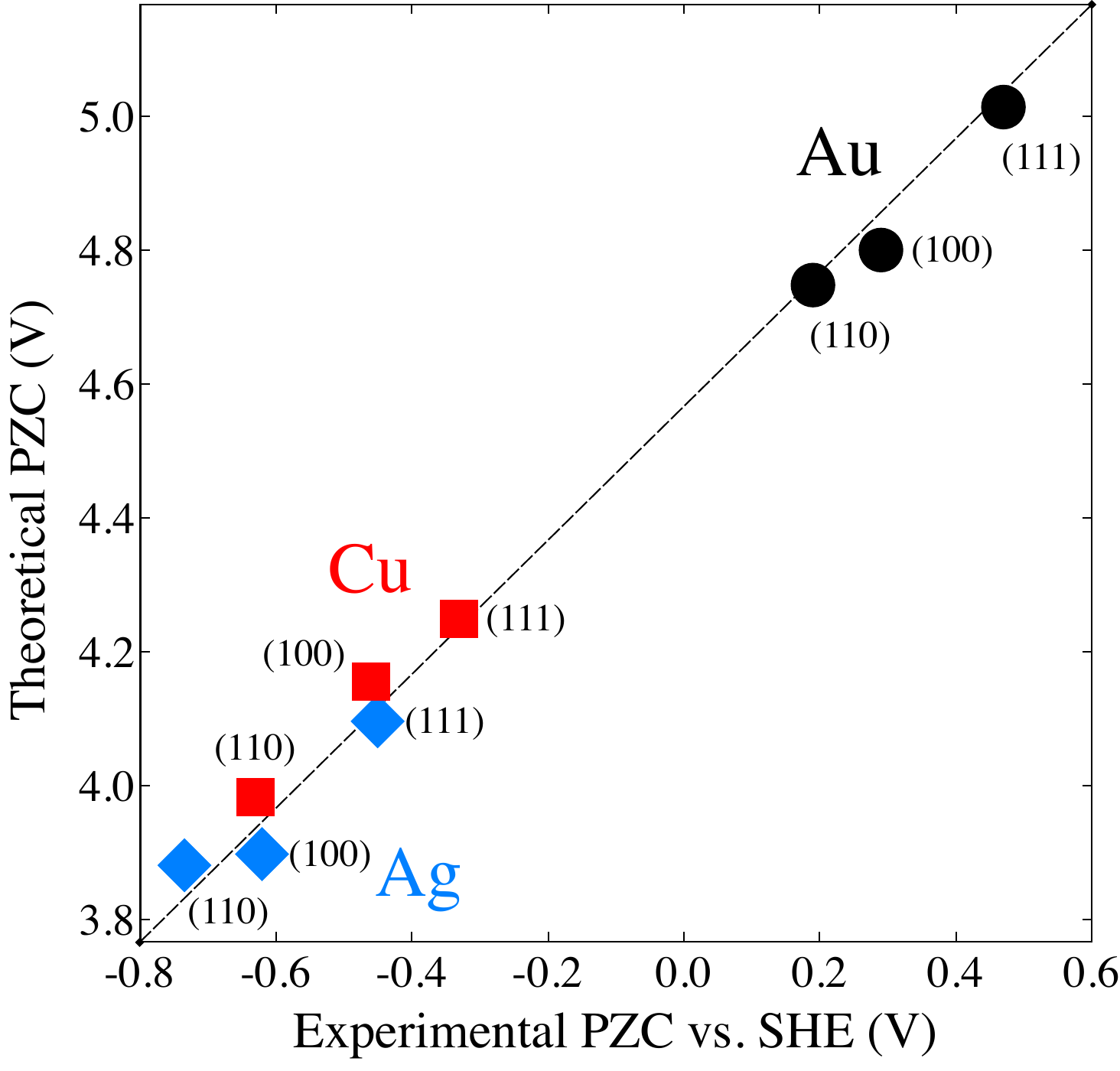}
  \caption{\label{fig:pzc} Comparison between the computed and the
    experimental potential of zero charge (PZC) with respect to the
    standard hydrogen electrode (SHE). The dashed line is a fit of
    $U^\mathrm{pred} = U^\mathrm{exp}+\Delta U_\mathrm{SHE}^\mathrm{pred}$
    to determine the theoretical potential of the SHE, $\Delta U_\mathrm{SHE}^\mathrm{pred}$. The
    experimental values are taken from
    Ref.~\onlinecite{Trasatti1999}.
  }
\end{figure}

\section{Validation}
\label{sec:pbz_validation}

We further validate the model by comparing with existing experimental and computational data. First, we compute the potential of zero charge (PZC) of various metallic slabs and compare them against the experimental values. Second, we compute the surface charge density of a Pt(111) slab as function of the applied external potential and compare the resulting value of the double layer capacitance with previous computations and experiments.

The DFT calculations for these two benchmarks and the following application are performed with VASP and the VASPsol module using the PAW formalism describing the electron-ion interactions and the PBE approximation for the exchange-correlation functional.\cite{VASP, paw, pbe} The Brillouin-zone integration employs an automatic mesh with 50 $k$-points per inverse \AA ngstrom with only one $k$-point in the direction perpendicular to the slab. 

We consider an electrolyte that consists of an aqueous solution of monovalent anions and cations of 1 M concentration. At room temperature this electrolyte has a relative permittivity of $\epsilon_\mathrm{b} = 78.4$ and a Debye length of $\lambda_\mathrm{D} = 3.04$~\AA. The parameters for the shape function, $\zeta(n(\vec r))$, are taken from Ref.~\onlinecite{Mathew2014, ta11} to be $n_\mathrm{c} = 0.0025$~\AA$^{-1}$ and $\sigma = 0.6$.

To properly resolve the interfacial region between the solute and the implicit electrolyte region and to obtain accurate values for the cavitation energy requires a high plane-wave basis set cutoff energy of 1000~eV. However, in our validation calculations and applications, we observe that the contribution of the cavitation energy to the solvation energies are negligibly small. We, therefore, set the effective surface tension parameter $\tau = 0$ for all following calculations. This also removes the requirement for such a high cutoff energy, and we find that a cutoff energy of 600~eV results in converged surface energies and PZC.

To change the applied potential, we adjust the electron count and then determine the corresponding potential from the shift in electrostatic potential as reflected in the shift in Fermi level. This takes advantage of the fact that the electrostatic potential goes to zero in the electrolyte region for the solution of the Poisson-Boltzmann equation, providing a reference for the electrostatic potential.

\begin{figure}[t]
  \includegraphics[width=\columnwidth]{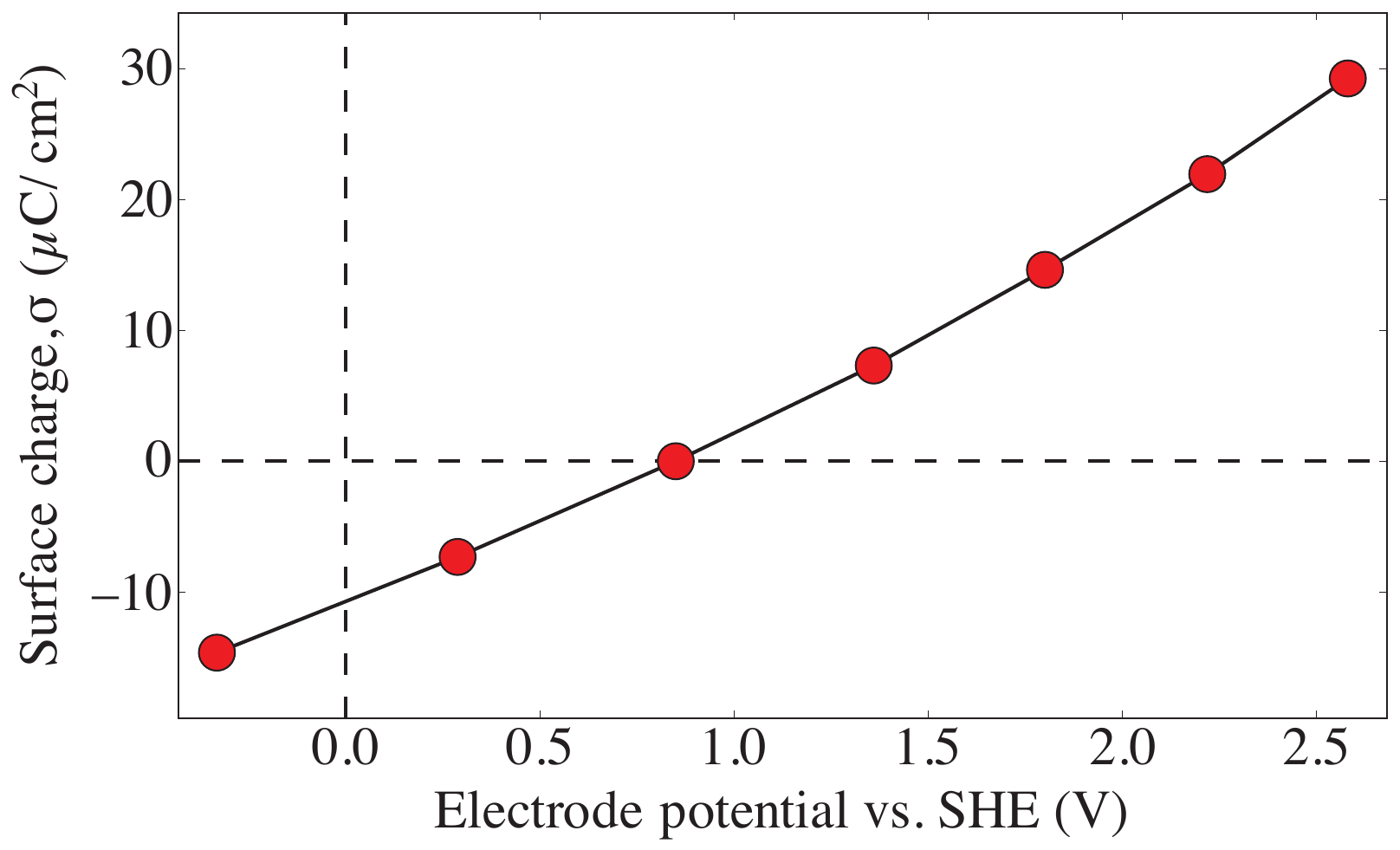} 
  \caption{\label{fig:pt_111} Surface charge density of the Pt(111)
    surface as a function of applied electrostatic potential computed
    with the implicit electrolyte module, VASPsol. The slope of the
    curve determines the capacitance of the dielectric double
    layer.}
\end{figure}

Figure~\ref{fig:pzc} compares the PZC for Cu, Ag, and Au surface facets calculated with the implicit electrolyte model as implemented in VASPsol with experimental data. The electrode PZC is defined as the electrostatic potential of a neutral metal electrode and is given by the Fermi energy with respect to a reference potential. A natural choice of reference for the implicit electrolyte model is the electrostatic potential in the bulk of the electrolyte, which is zero in any solution of the Poisson-Boltzmann equation. In experiments, the reference is frequently chosen as the standard-hydrogen electrode (SHE). The dashed line in Fig.~\ref{fig:pzc} presents the fit of $U^\mathrm{pred} = U^\mathrm{exp}+\Delta U_\mathrm{SHE}^\mathrm{pred}$ to obtain the shift, $\Delta U_\mathrm{SHE}^\mathrm{pred}$, between the experimental and the computed PZC, {\it i.e.} the potential of the SHE for our electrolyte model. The resulting $\Delta U_\mathrm{SHE}^\mathrm{pred}=4.6$~V compares well with the previously reported computed value of 4.7~V for a similar solvation model.\cite{ta11}

Figure~\ref{fig:pt_111} shows the calculated surface charge density, $\sigma$, of a Pt(111) surface as a function of applied electrostatic potential, $U$. We find that the PZC for the Pt(111) electrode is 0.85~V, in good agreement with previous computational studies\cite{ta4, ta11}. However, this value is at the high end of the experimentally reported results.\cite{Pajkossy2001, Trasatti1999} This discrepancy is likely due to adsorption on Pt(111). The work by Sekong and Gro\ss\ shows that Pt(111) is covered at potentials below 0.5~V by hydrogen, followed by the so-called double layer region between 0.5 and 0.75~V with no adsorbates present and OH adsorption at more positive potentials.~\cite{vsol_sakong} The presence of adsorbates in the experiments alters the PZC and precludes a direct comparison with our calculations.

The linear slope of $\sigma(U)$ is the double-layer capacitance. Fitting a quadratic polynomial to the data and evaluating the slope at the pzc yields a double-layer capacitance for the Pt(111) surface at 1 M concentration of 14~$\mu$F/cm$^2$. This agrees perfectly with the previously reported computed result\cite{ta11} and is close to the experimental value of 20~$\mu$F/cm$^2$.\cite{Pajkossy2001}

\begin{figure}[t]
  \includegraphics[width=\columnwidth]{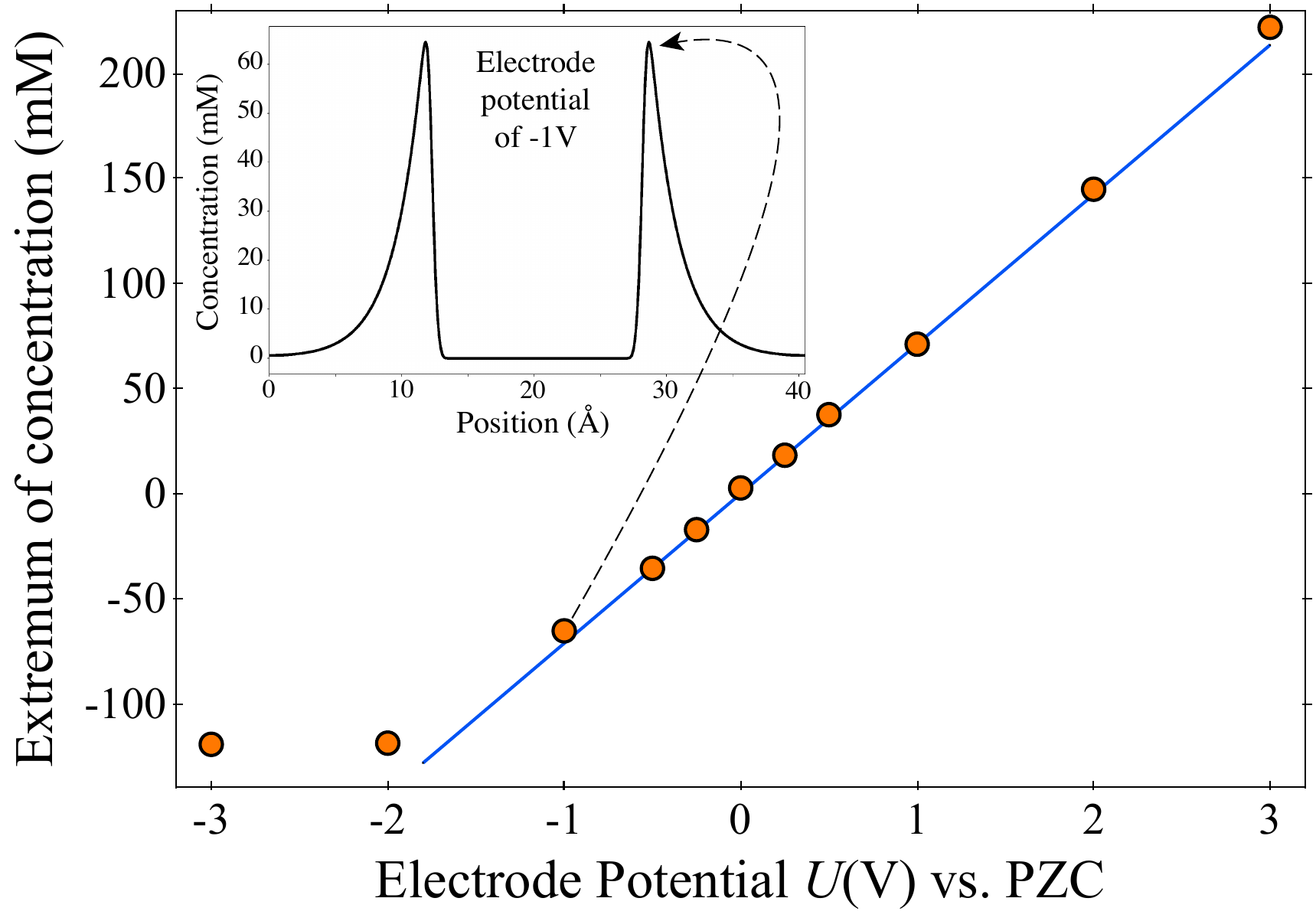}
  \caption{\label{fig:Cu_Ionic} The extrema of the ion concentrations over the potential range of $-3$ to +3~V relative to the PZC in an electrolyte with ion concentration of 2~M. The negative and positive concentrations correspond to excess of cations and anions, respectively, in the double layer. The solid blue line is a linear fit between $-1$ and +2~V, indicating the large linear potential range. The inset shows the $xy$ averaged ion concentration profile along the $z$-direction at a potential of $-1$~V; the central region corresponds to the solute slab with zero ion concentration.
  }
\end{figure}

Our implementation of the electrolyte model neglects the finite volume of the ions in the electrolyte. Hence, at larger applied potentials, the model may predict unphysically large ion concentrations to compensate the surface charge. To understand how this approximation limits the applied potential, we perform calculations on a Cu(111) slab in an electrolyte with 2~M concentration for a range of applied potentials from $-3$ to +3~V.

Figure~\ref{fig:Cu_Ionic} shows the extrema of the net ion concentration in the double-layer region as a function of the applied potential (written by VASPsol into the file RHOION). The negative and positive concentrations correspond to excess cations and anions near the surface, respectively. For a large range of potentials from $-2$ to +3~V, the net ion concentration depends nearly linear on the applied potential. At negative potentials below $-2$~V, we observe that electrons leak from the slab into the electrolyte region, rendering the results of the electrolyte model unphysical. Hence, care should be taken when utilizing the model at large negative potentials.

The inset in Fig.~\ref{fig:Cu_Ionic} illustrates the average net ion concentration along the direction perpendicular to the Cu(111) surface for an applied potential of $-1$~V. The ion concentration decays exponentially away from the solid-liquid interface and approaches zero in the bulk electrolyte. We selected an electrolyte region of 30~{\AA} to illustrate the need to converge the results with respect to the size of the electrolyte domain. Here, the size is not quite sufficient to reach a negligible ion concentration. While this does not affect the considered extrema of the ion concentration, care must be taken to ensure convergence of the results with the size of the electrolyte region. As general guidance, an electrolyte region $>10 \lambda$ should suffice.\cite{Steinmann2016a}

At a large applied potential of +3~V, we find a net ion concentration of 0.2~M, which is 10\% of the electrolyte concentration, far from the 50\% excess that would introduce a 1\% error as discussed following Eq. \ref{eq:linPBeq}. Here, the concentrations of anions and cations are 2.1 and 1.9~M, respectively, and the error introduced by the linear approximation to the sinh function is less than 0.1\%. 
Thus, even for a considerable potential of +3~V, the linear electrolyte model does not result in unphysically large ion concentrations.

\begin{figure}[t]
  \includegraphics[width=\columnwidth]{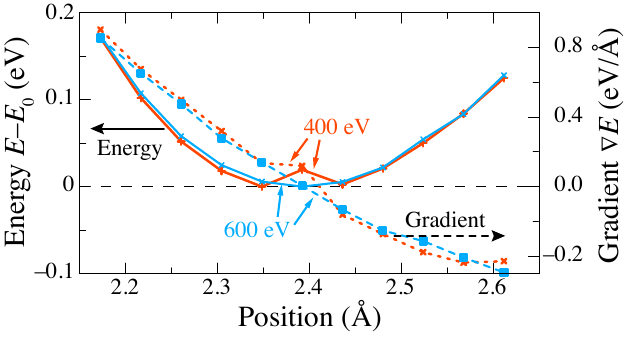}
  \caption{\label{fig:Gradient} Analytical gradient and relative energy as a function of the distance of a Na ion from the Au(111) surface, at a potential of about $-1.9$~V vs.\ the potential of the SHE. The plane-wave cutoff of 400~eV leads to numerical noise, which disappears at 600~eV.}
\end{figure}

Finally, we validate the accuracy of the implemented analytic force evaluation. As an example, we have chosen the adsorption of a Na atom on the Au(111) surface. The surface charge was set to 0.4 $e^-$ for the symmetric p($2\times2$) surface, corresponding to an electrochemical potential of about $-1.9$~V vs. the SHE, while the neutral Na@Au(111) surface corresponds to a potential of $-2.6$~V. Fig.~\ref{fig:Gradient} shows that the agreement between the analytical gradient and the minimum for the energy is excellent, provided a sufficiently high cutoff energy of 600~eV is used.

\begin{figure}[t]
  \includegraphics[width=\columnwidth]{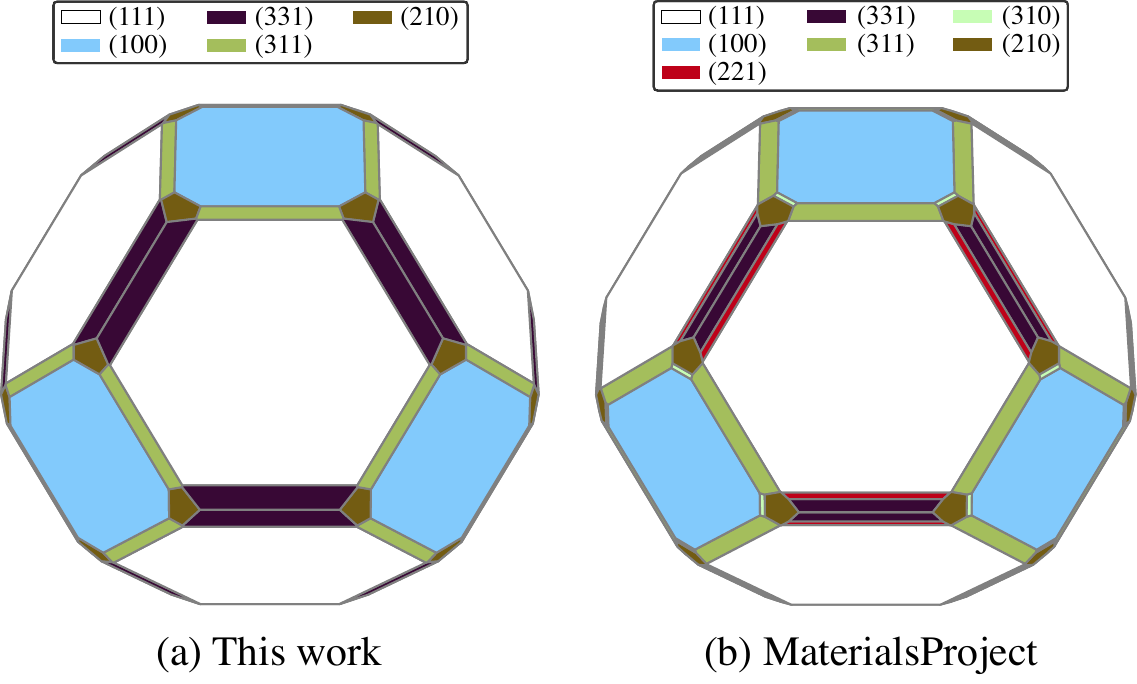}
  \caption{\label{fig:Cu_vacuumplots} Comparison of the predicted shape of a Cu crystal in vacuum (a) for this work obtained by the Wulff construction using the surface energies of the (100), (110), (111), (210), (221), (311), and (331) facets compared with (b) the  MaterialsProject database.\cite{MP_wulff}
  }
\end{figure}

\section{Crystal Shape Control by Applied Potential}
\label{sec:pbz_applications}

The implicit electrolyte model, VASPsol, enables the study of electrode surfaces in realistic environments and under various conditions, such as of an electrode immersed in an electrolyte with an external potential applied. To illustrate the utility of the solvation model, we determine how an applied potential changes the equilibrium shape of a Cu crystal in a 1 M electrolyte. The crystal shape is controlled by the surface energies. We show that the surface energies are sensitive to the applied potential and that each facet follows different trends, leading to opportunities for the practical control of the shape of nanocrystals.

Using the MPInterfaces framework,\cite{Mathew2016mpinterfaces} we construct slabs of minimum 10~\AA\ thickness for the(111), (100), (110), (210), (221), (311) and (331) facets of Cu. The different facets for Cu are chosen based on the crystal facets included in the MaterialsProject database.\cite{MP_wulff} We calculate the surface energy of each slab in vacuum and electrolyte by relaxing the top and bottom layers and keeping the middle layers of atoms fixed to their bulk positions. We employ a vacuum spacing of 30~{\AA} to ensure that the electrostatic interactions between periodic images of the slabs are negligibly small. From the surface energies, we determine the resulting shape using the Wulff construction as implemented in the pymatgen framework.\cite{MP_wulff}

\begin{figure}[t]
  \includegraphics[width=\columnwidth]{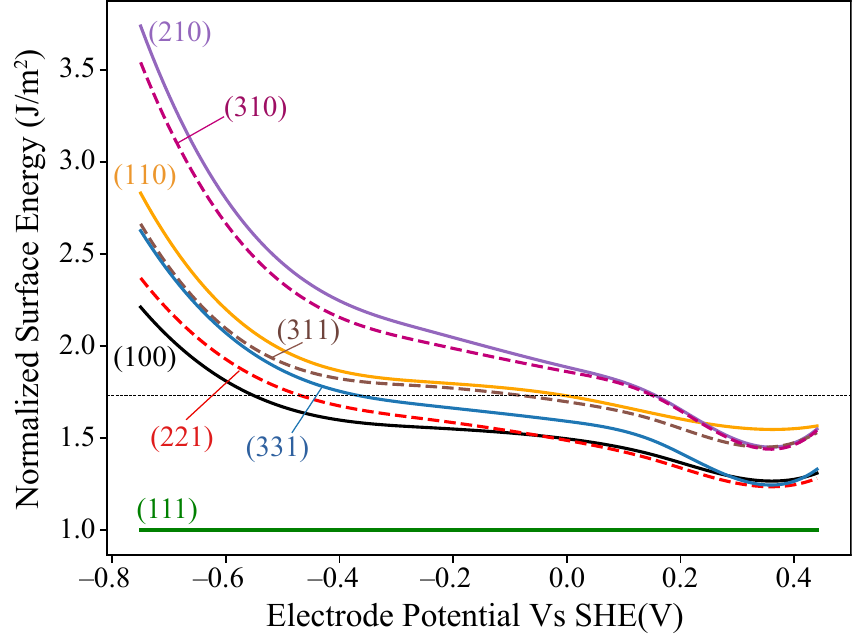}
  \caption{\label{fig:Cu_relaxedenergies} The surface energies, $\gamma_{hkl}$, of the various facets of Cu with respect to surface energies, $\gamma_{111}$, of the (111) facet as a function of applied potential. The dashed horizontal line indicates the ratio of the (100) and (111) surface energies, $\gamma_{100}/\gamma_{111}$ equals $\sqrt{3}$, which corresponds to a change in shape from an octahedron to a truncated octahedron.}
\end{figure}

Figure~\ref{fig:Cu_vacuumplots} compares the crystal shape of Cu in vacuum obtained in our calculations with MaterialsProject. The MaterialsProject calculations used a cutoff energy of 400~eV and a vacuum spacing of 10~\AA\, which both are lower than the cutoff energy of 600~eV and the vacuum spacing of 30~{\AA} used in this work. We find that our predicted surface energies agree with MaterialsProject within 2\% for all the facets. However, we note that even such small deviations in surface energies matter for the details of the crystal shape, where the high index (310) and (221) facets do not show up in our calculation of the equilibrium crystal shape of Cu. These small but observable deviations demonstrate that the crystal shape is sensitive to small variations in the surface energies.

For the calculation of the Cu surface energies in the 1 M electrolyte, we vary the number of electrons in the slabs and determine the change in the corresponding applied electrostatic potential as described in Sec.~\ref{sec:implementation}. We calculate the surface energies for each of the above facets in vacuum over a range of applied potentials of approximately $\pm$1~eV around the corresponding PZC. We then perform a spline interpolation to obtain the surface energies as a function of the applied potential. This transformation enables the comparison of the surface energies for a given applied potential and is made possible by the absolute reference potential provided by the solvation model. From the results we obtain the equilibrium crystal shape as a function of applied potentials by the Wulff construction.

For copper, the (111) facets exhibit the lowest surface energy over the whole potential range. Figure~\ref{fig:Cu_relaxedenergies} shows how the ratios of the surface energies of the copper (hkl) facets relative to the (111) facet vary as a function of applied potential. As expected, the surface energies of all facets systematically increase with potential, however, their ratios relative to the (111) facet decreases. Due to this, the crystal shape changes as a function of applied potential.

\begin{figure}[t]
  \includegraphics[width=\columnwidth]{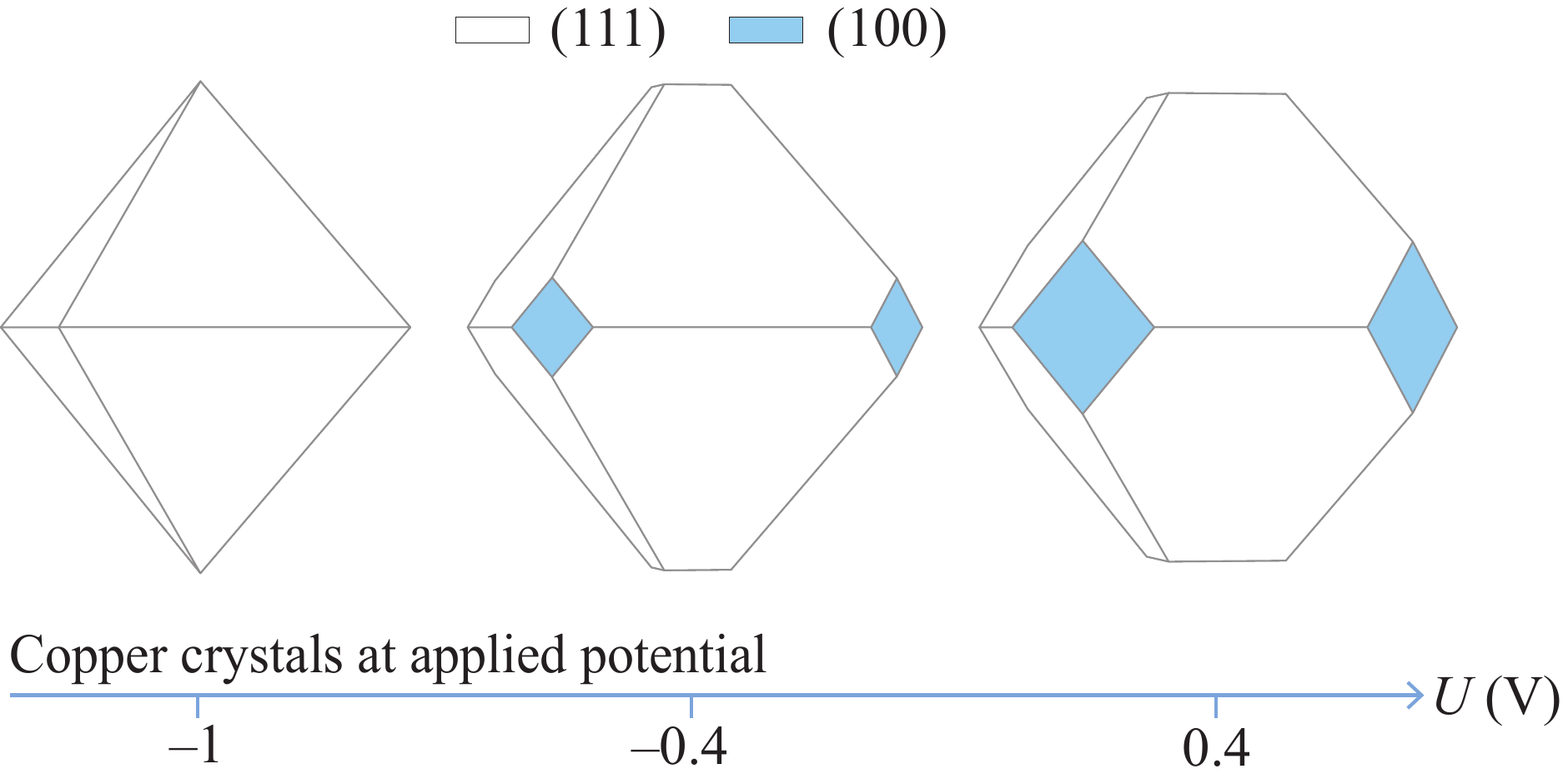}
  \caption{\label{fig:Wulff plots} Variation of the equilibrium crystal shape of Cu with applied electrode potential. The equilibrium shape transitions from an octahedron at negative potentials to a truncated octahedron upon increasing applied potentials.}
\end{figure}

Figure~\ref{fig:Wulff plots} illustrates the change in crystal shape with increasing applied potential. At negative potentials,  the (111) facets dominate and lead to an octahedral shape. With increasing potential, the ratio of the (100) to  (111) surface energies decreases sufficiently to lead to a change in shape to a truncated octahedron. The transition occurs around a potential of $-0.56$~V for Cu. With further increase of potential, the area of the (100) facets increases. Even though the ratio of the surface energies of other facets are also lowered, the change is not sufficient for other facet to show up in the equilibrium crystal shape.

The prediction for the changes in the equilibrium shape of Cu crystal as a function of applied electric potential demonstrate the utility of the VASPsol solvation model for computational electrochemistry simulations using density-functional theory and provides the opportunity to design nanocrystals of specific shape through the application of an electrostatic potential during growth at an applied potential.

\section{Conclusions}
\label{sec:pbz_conclusions}

Solid/liquid interfaces between electrodes and electrolytes in electrochemical cells present a complex system that benefits from insights provided by computational studies to supplement and explain experimental observations. We developed and implemented a self-consistent computational framework that provides an efficient implicit description of electrode/electrolyte interfaces within density-functional theory through the solution of the linearized Poisson-Boltzmann equation. The electrolyte model is implemented in the widely used DFT code VASP, and the implementation is made available as a free and open-source module, VASPsol, hosted on GitHub at \url{https://github.com/henniggroup/VASPsol}. We showed that this model enables ab-initio electrochemistry studies at the DFT level. We validated the model by comparing the calculated electrode potentials of zero charge for various metal electrodes with experimental values and the calculated double-layer capacitance of Pt(111) in a 1 M solution to previous calculations and experiments. We tested the validity of the linear approximation at large applied potentials and the accuracy of the analytic expressions for the forces. To illustrate the usefulness of the model and the implementation, we apply the model to determine the equilibrium shape of Cu crystal as a function of applied electrostatic potential. We predict a change in shape from an octahedron to a truncated octahedron with increasing potential. These calculations illustrate the utility of the implicit VASPsol model for computational electrochemistry simulations.

\begin{acknowledgments}
  This work was supported by the National Science Foundation under awards Nos.\ DMR-1542776, ACI-1440547, and OAC-1740251. This research used computational resources of the Texas Advanced Computing Center under Contract Number TG-DMR050028N and of the University of Florida Research Computing Center. Part of this research was performed while the authors were visiting the Institute for Pure and Applied Mathematics (IPAM), which is supported by the National Science Foundation (NSF).  \end{acknowledgments}

\bibliography{Manuscript}

\begin{thebibliography}{40}%
\makeatletter
\providecommand \@ifxundefined [1]{%
 \@ifx{#1\undefined}
}%
\providecommand \@ifnum [1]{%
 \ifnum #1\expandafter \@firstoftwo
 \else \expandafter \@secondoftwo
 \fi
}%
\providecommand \@ifx [1]{%
 \ifx #1\expandafter \@firstoftwo
 \else \expandafter \@secondoftwo
 \fi
}%
\providecommand \natexlab [1]{#1}%
\providecommand \enquote  [1]{``#1''}%
\providecommand \bibnamefont  [1]{#1}%
\providecommand \bibfnamefont [1]{#1}%
\providecommand \citenamefont [1]{#1}%
\providecommand \href@noop [0]{\@secondoftwo}%
\providecommand \href [0]{\begingroup \@sanitize@url \@href}%
\providecommand \@href[1]{\@@startlink{#1}\@@href}%
\providecommand \@@href[1]{\endgroup#1\@@endlink}%
\providecommand \@sanitize@url [0]{\catcode `\\12\catcode `\$12\catcode
  `\&12\catcode `\#12\catcode `\^12\catcode `\_12\catcode `\%12\relax}%
\providecommand \@@startlink[1]{}%
\providecommand \@@endlink[0]{}%
\providecommand \url  [0]{\begingroup\@sanitize@url \@url }%
\providecommand \@url [1]{\endgroup\@href {#1}{\urlprefix }}%
\providecommand \urlprefix  [0]{URL }%
\providecommand \Eprint [0]{\href }%
\providecommand \doibase [0]{http://dx.doi.org/}%
\providecommand \selectlanguage [0]{\@gobble}%
\providecommand \bibinfo  [0]{\@secondoftwo}%
\providecommand \bibfield  [0]{\@secondoftwo}%
\providecommand \translation [1]{[#1]}%
\providecommand \BibitemOpen [0]{}%
\providecommand \bibitemStop [0]{}%
\providecommand \bibitemNoStop [0]{.\EOS\space}%
\providecommand \EOS [0]{\spacefactor3000\relax}%
\providecommand \BibitemShut  [1]{\csname bibitem#1\endcsname}%
\let\auto@bib@innerbib\@empty
\bibitem [{\citenamefont {Wang}\ \emph {et~al.}(2010)\citenamefont {Wang},
  \citenamefont {Xin}, \citenamefont {Yu}, \citenamefont {Wang}, \citenamefont
  {Rus}, \citenamefont {Muller},\ and\ \citenamefont {Abru\~na}}]{abruna}%
  \BibitemOpen
  \bibfield  {author} {\bibinfo {author} {\bibfnamefont {D.}~\bibnamefont
  {Wang}}, \bibinfo {author} {\bibfnamefont {H.~L.}\ \bibnamefont {Xin}},
  \bibinfo {author} {\bibfnamefont {Y.}~\bibnamefont {Yu}}, \bibinfo {author}
  {\bibfnamefont {H.}~\bibnamefont {Wang}}, \bibinfo {author} {\bibfnamefont
  {E.}~\bibnamefont {Rus}}, \bibinfo {author} {\bibfnamefont {D.~A.}\
  \bibnamefont {Muller}}, \ and\ \bibinfo {author} {\bibfnamefont {H.~D.}\
  \bibnamefont {Abru\~na}},\ }\href {\doibase 10.1021/ja107874u} {\bibfield
  {journal} {\bibinfo  {journal} {J. Am. Chem. Soc.}\ }\textbf {\bibinfo
  {volume} {132}},\ \bibinfo {pages} {17664} (\bibinfo {year}
  {2010})}\BibitemShut {NoStop}%
\bibitem [{\citenamefont {Ha}, \citenamefont {Islam},\ and\ \citenamefont
  {Robinson}(2012)}]{robinson}%
  \BibitemOpen
  \bibfield  {author} {\bibinfo {author} {\bibfnamefont {D.-H.}\ \bibnamefont
  {Ha}}, \bibinfo {author} {\bibfnamefont {M.~A.}\ \bibnamefont {Islam}}, \
  and\ \bibinfo {author} {\bibfnamefont {R.~D.}\ \bibnamefont {Robinson}},\
  }\href {\doibase 10.1021/nl3019559} {\bibfield  {journal} {\bibinfo
  {journal} {Nano Lett.}\ }\textbf {\bibinfo {volume} {12}},\ \bibinfo {pages}
  {5122} (\bibinfo {year} {2012})}\BibitemShut {NoStop}%
\bibitem [{\citenamefont {Liu}, \citenamefont {Luais},\ and\ \citenamefont
  {Gooding}(2011)}]{npelectrochem}%
  \BibitemOpen
  \bibfield  {author} {\bibinfo {author} {\bibfnamefont {G.}~\bibnamefont
  {Liu}}, \bibinfo {author} {\bibfnamefont {E.}~\bibnamefont {Luais}}, \ and\
  \bibinfo {author} {\bibfnamefont {J.~J.}\ \bibnamefont {Gooding}},\ }\href
  {\doibase 10.1021/la104373v} {\bibfield  {journal} {\bibinfo  {journal}
  {Langmuir}\ }\textbf {\bibinfo {volume} {27}},\ \bibinfo {pages} {4176}
  (\bibinfo {year} {2011})}\BibitemShut {NoStop}%
\bibitem [{\citenamefont {Choi}\ \emph {et~al.}(2011)\citenamefont {Choi},
  \citenamefont {Bealing}, \citenamefont {Bian}, \citenamefont {Hughes},
  \citenamefont {Zhang}, \citenamefont {Smilgies}, \citenamefont {Hennig},
  \citenamefont {Engstrom},\ and\ \citenamefont {Hanrath}}]{hanrath}%
  \BibitemOpen
  \bibfield  {author} {\bibinfo {author} {\bibfnamefont {J.~J.}\ \bibnamefont
  {Choi}}, \bibinfo {author} {\bibfnamefont {C.~R.}\ \bibnamefont {Bealing}},
  \bibinfo {author} {\bibfnamefont {K.}~\bibnamefont {Bian}}, \bibinfo {author}
  {\bibfnamefont {K.~J.}\ \bibnamefont {Hughes}}, \bibinfo {author}
  {\bibfnamefont {W.}~\bibnamefont {Zhang}}, \bibinfo {author} {\bibfnamefont
  {D.-M.}\ \bibnamefont {Smilgies}}, \bibinfo {author} {\bibfnamefont {R.~G.}\
  \bibnamefont {Hennig}}, \bibinfo {author} {\bibfnamefont {J.~R.}\
  \bibnamefont {Engstrom}}, \ and\ \bibinfo {author} {\bibfnamefont
  {T.}~\bibnamefont {Hanrath}},\ }\href {\doibase 10.1021/ja110454b} {\bibfield
   {journal} {\bibinfo  {journal} {J. Am. Chem. Soc.}\ }\textbf {\bibinfo
  {volume} {133}},\ \bibinfo {pages} {3131} (\bibinfo {year}
  {2011})}\BibitemShut {NoStop}%
\bibitem [{\citenamefont {Moyano}\ and\ \citenamefont
  {Rotello}(2011)}]{npinterface}%
  \BibitemOpen
  \bibfield  {author} {\bibinfo {author} {\bibfnamefont {D.~F.}\ \bibnamefont
  {Moyano}}\ and\ \bibinfo {author} {\bibfnamefont {V.~M.}\ \bibnamefont
  {Rotello}},\ }\href {\doibase 10.1021/la2004535} {\bibfield  {journal}
  {\bibinfo  {journal} {Langmuir}\ }\textbf {\bibinfo {volume} {27}},\ \bibinfo
  {pages} {10376} (\bibinfo {year} {2011})}\BibitemShut {NoStop}%
\bibitem [{\citenamefont {Liu}\ \emph {et~al.}(2012)\citenamefont {Liu},
  \citenamefont {Chen}, \citenamefont {Yakovenko},\ and\ \citenamefont
  {Zhou}}]{liu2012interconversion}%
  \BibitemOpen
  \bibfield  {author} {\bibinfo {author} {\bibfnamefont {T.-F.}\ \bibnamefont
  {Liu}}, \bibinfo {author} {\bibfnamefont {Y.-P.}\ \bibnamefont {Chen}},
  \bibinfo {author} {\bibfnamefont {A.~A.}\ \bibnamefont {Yakovenko}}, \ and\
  \bibinfo {author} {\bibfnamefont {H.-C.}\ \bibnamefont {Zhou}},\ }\href@noop
  {} {\bibfield  {journal} {\bibinfo  {journal} {J. Am. Chem. Soc.}\ }\textbf
  {\bibinfo {volume} {134}},\ \bibinfo {pages} {17358} (\bibinfo {year}
  {2012})}\BibitemShut {NoStop}%
\bibitem [{\citenamefont {Agubra}\ and\ \citenamefont
  {Fergus}(2014)}]{agubra2014formation}%
  \BibitemOpen
  \bibfield  {author} {\bibinfo {author} {\bibfnamefont {V.~A.}\ \bibnamefont
  {Agubra}}\ and\ \bibinfo {author} {\bibfnamefont {J.~W.}\ \bibnamefont
  {Fergus}},\ }\href@noop {} {\bibfield  {journal} {\bibinfo  {journal} {J.
  Power Sources}\ }\textbf {\bibinfo {volume} {268}},\ \bibinfo {pages} {153}
  (\bibinfo {year} {2014})}\BibitemShut {NoStop}%
\bibitem [{\citenamefont {Ismail-Beigi}\ and\ \citenamefont
  {Arias}(2000)}]{ta2}%
  \BibitemOpen
  \bibfield  {author} {\bibinfo {author} {\bibfnamefont {S.}~\bibnamefont
  {Ismail-Beigi}}\ and\ \bibinfo {author} {\bibfnamefont {T.~A.}\ \bibnamefont
  {Arias}},\ }\href@noop {} {\bibfield  {journal} {\bibinfo  {journal} {Comp.
  Phys. Comm.}\ }\textbf {\bibinfo {volume} {128}},\ \bibinfo {pages} {1}
  (\bibinfo {year} {2000})}\BibitemShut {NoStop}%
\bibitem [{\citenamefont {White}\ \emph {et~al.}(2000)\citenamefont {White},
  \citenamefont {Schwegler}, \citenamefont {Galli},\ and\ \citenamefont
  {Gygi}}]{Galli2000}%
  \BibitemOpen
  \bibfield  {author} {\bibinfo {author} {\bibfnamefont {J.~A.}\ \bibnamefont
  {White}}, \bibinfo {author} {\bibfnamefont {E.}~\bibnamefont {Schwegler}},
  \bibinfo {author} {\bibfnamefont {G.}~\bibnamefont {Galli}}, \ and\ \bibinfo
  {author} {\bibfnamefont {F.}~\bibnamefont {Gygi}},\ }\href@noop {} {\bibfield
   {journal} {\bibinfo  {journal} {J. Chem. Phys.}\ }\textbf {\bibinfo {volume}
  {113}} (\bibinfo {year} {2000})}\BibitemShut {NoStop}%
\bibitem [{\citenamefont {Petrosyan}, \citenamefont {Rigos},\ and\
  \citenamefont {Arias}(2005)}]{ta5}%
  \BibitemOpen
  \bibfield  {author} {\bibinfo {author} {\bibfnamefont {S.~A.}\ \bibnamefont
  {Petrosyan}}, \bibinfo {author} {\bibfnamefont {A.~A.}\ \bibnamefont
  {Rigos}}, \ and\ \bibinfo {author} {\bibfnamefont {T.~A.}\ \bibnamefont
  {Arias}},\ }\href@noop {} {\bibfield  {journal} {\bibinfo  {journal} {J.
  Phys. Chem. B}\ }\textbf {\bibinfo {volume} {109}},\ \bibinfo {pages} {15436}
  (\bibinfo {year} {2005})}\BibitemShut {NoStop}%
\bibitem [{\citenamefont {Cramer}\ and\ \citenamefont
  {Truhlar}(2008)}]{cramer}%
  \BibitemOpen
  \bibfield  {author} {\bibinfo {author} {\bibfnamefont {C.~J.}\ \bibnamefont
  {Cramer}}\ and\ \bibinfo {author} {\bibfnamefont {D.~G.}\ \bibnamefont
  {Truhlar}},\ }\href {\doibase 10.1021/ar800019z} {\bibfield  {journal}
  {\bibinfo  {journal} {Acc. Chem. Res.}\ }\textbf {\bibinfo {volume} {41}},\
  \bibinfo {pages} {760} (\bibinfo {year} {2008})}\BibitemShut {NoStop}%
\bibitem [{\citenamefont {Lischner}\ and\ \citenamefont {Arias}(2010)}]{ta3}%
  \BibitemOpen
  \bibfield  {author} {\bibinfo {author} {\bibfnamefont {J.}~\bibnamefont
  {Lischner}}\ and\ \bibinfo {author} {\bibfnamefont {T.~A.}\ \bibnamefont
  {Arias}},\ }\href@noop {} {\bibfield  {journal} {\bibinfo  {journal} {J .
  Chem. Phys. B}\ }\textbf {\bibinfo {volume} {114}},\ \bibinfo {pages} {1946}
  (\bibinfo {year} {2010})}\BibitemShut {NoStop}%
\bibitem [{\citenamefont {Andreussi}, \citenamefont {Dabo},\ and\ \citenamefont
  {Marzari}(2012)}]{Marzari}%
  \BibitemOpen
  \bibfield  {author} {\bibinfo {author} {\bibfnamefont {O.}~\bibnamefont
  {Andreussi}}, \bibinfo {author} {\bibfnamefont {I.}~\bibnamefont {Dabo}}, \
  and\ \bibinfo {author} {\bibfnamefont {N.}~\bibnamefont {Marzari}},\
  }\href@noop {} {\bibfield  {journal} {\bibinfo  {journal} {J. Chem. Phys}\
  }\textbf {\bibinfo {volume} {136}},\ \bibinfo {pages} {064102} (\bibinfo
  {year} {2012})}\BibitemShut {NoStop}%
\bibitem [{\citenamefont {Ringe}\ \emph {et~al.}(2016)\citenamefont {Ringe},
  \citenamefont {Oberhofer}, \citenamefont {Hille}, \citenamefont {Matera},\
  and\ \citenamefont {Reuter}}]{Ringe2016}%
  \BibitemOpen
  \bibfield  {author} {\bibinfo {author} {\bibfnamefont {S.}~\bibnamefont
  {Ringe}}, \bibinfo {author} {\bibfnamefont {H.}~\bibnamefont {Oberhofer}},
  \bibinfo {author} {\bibfnamefont {C.}~\bibnamefont {Hille}}, \bibinfo
  {author} {\bibfnamefont {S.}~\bibnamefont {Matera}}, \ and\ \bibinfo {author}
  {\bibfnamefont {K.}~\bibnamefont {Reuter}},\ }\href@noop {} {\bibfield
  {journal} {\bibinfo  {journal} {J. Chem. Theory Comput.}\ }\textbf {\bibinfo
  {volume} {12}},\ \bibinfo {pages} {4052} (\bibinfo {year}
  {2016})}\BibitemShut {NoStop}%
\bibitem [{\citenamefont {Mathew}\ \emph {et~al.}(2014)\citenamefont {Mathew},
  \citenamefont {Sundararaman}, \citenamefont {Letchworth-Weaver},
  \citenamefont {Arias},\ and\ \citenamefont {Hennig}}]{Mathew2014}%
  \BibitemOpen
  \bibfield  {author} {\bibinfo {author} {\bibfnamefont {K.}~\bibnamefont
  {Mathew}}, \bibinfo {author} {\bibfnamefont {R.}~\bibnamefont
  {Sundararaman}}, \bibinfo {author} {\bibfnamefont {K.}~\bibnamefont
  {Letchworth-Weaver}}, \bibinfo {author} {\bibfnamefont {T.~A.}\ \bibnamefont
  {Arias}}, \ and\ \bibinfo {author} {\bibfnamefont {R.~G.}\ \bibnamefont
  {Hennig}},\ }\href {\doibase http://dx.doi.org/10.1063/1.4865107} {\bibfield
  {journal} {\bibinfo  {journal} {J. Chem. Phys.}\ }\textbf {\bibinfo {volume}
  {140}},\ \bibinfo {eid} {084106} (\bibinfo {year} {2014})}\BibitemShut
  {NoStop}%
\bibitem [{\citenamefont {Mathew}\ and\ \citenamefont
  {Hennig}(2015)}]{VASPsol_GitHub}%
  \BibitemOpen
  \bibfield  {author} {\bibinfo {author} {\bibfnamefont {K.}~\bibnamefont
  {Mathew}}\ and\ \bibinfo {author} {\bibfnamefont {R.~G.}\ \bibnamefont
  {Hennig}},\ }\href@noop {} {\enquote {\bibinfo {title} {{VASPsol} -
  {S}olvation model for the plane wave {DFT} code {VASP}},}\ }\bibinfo
  {howpublished} {\url{https://github.com/henniggroup/VASPsol}} (\bibinfo
  {year} {2015})\BibitemShut {NoStop}%
\bibitem [{\citenamefont {Kresse}\ and\ \citenamefont
  {Furthm\"uller}(1996)}]{VASP}%
  \BibitemOpen
  \bibfield  {author} {\bibinfo {author} {\bibfnamefont {G.}~\bibnamefont
  {Kresse}}\ and\ \bibinfo {author} {\bibfnamefont {J.}~\bibnamefont
  {Furthm\"uller}},\ }\href@noop {} {\bibfield  {journal} {\bibinfo  {journal}
  {Comput. Mater. Sci.}\ }\textbf {\bibinfo {volume} {6}},\ \bibinfo {pages}
  {15} (\bibinfo {year} {1996})}\BibitemShut {NoStop}%
\bibitem [{\citenamefont {Michel}\ \emph {et~al.}(2014)\citenamefont {Michel},
  \citenamefont {Zaffran}, \citenamefont {Ruppert}, \citenamefont
  {Matras-Michalska}, \citenamefont {Jedrzejczyk}, \citenamefont {Grams},\ and\
  \citenamefont {Sautet}}]{vsol_ketone}%
  \BibitemOpen
  \bibfield  {author} {\bibinfo {author} {\bibfnamefont {C.}~\bibnamefont
  {Michel}}, \bibinfo {author} {\bibfnamefont {J.}~\bibnamefont {Zaffran}},
  \bibinfo {author} {\bibfnamefont {A.~M.}\ \bibnamefont {Ruppert}}, \bibinfo
  {author} {\bibfnamefont {J.}~\bibnamefont {Matras-Michalska}}, \bibinfo
  {author} {\bibfnamefont {M.}~\bibnamefont {Jedrzejczyk}}, \bibinfo {author}
  {\bibfnamefont {J.}~\bibnamefont {Grams}}, \ and\ \bibinfo {author}
  {\bibfnamefont {P.}~\bibnamefont {Sautet}},\ }\href@noop {} {\bibfield
  {journal} {\bibinfo  {journal} {Chem. Commun.}\ }\textbf {\bibinfo {volume}
  {50}},\ \bibinfo {pages} {12450} (\bibinfo {year} {2014})}\BibitemShut
  {NoStop}%
\bibitem [{\citenamefont {Kim}\ \emph {et~al.}(2016)\citenamefont {Kim},
  \citenamefont {Park}, \citenamefont {Lee}, \citenamefont {Seong},
  \citenamefont {Lim}, \citenamefont {Bae}, \citenamefont {Kim}, \citenamefont
  {Kim}, \citenamefont {Ryu},\ and\ \citenamefont {Kang}}]{vsol_jinsoo}%
  \BibitemOpen
  \bibfield  {author} {\bibinfo {author} {\bibfnamefont {J.}~\bibnamefont
  {Kim}}, \bibinfo {author} {\bibfnamefont {H.}~\bibnamefont {Park}}, \bibinfo
  {author} {\bibfnamefont {B.}~\bibnamefont {Lee}}, \bibinfo {author}
  {\bibfnamefont {W.~M.}\ \bibnamefont {Seong}}, \bibinfo {author}
  {\bibfnamefont {H.-D.}\ \bibnamefont {Lim}}, \bibinfo {author} {\bibfnamefont
  {Y.}~\bibnamefont {Bae}}, \bibinfo {author} {\bibfnamefont {H.}~\bibnamefont
  {Kim}}, \bibinfo {author} {\bibfnamefont {W.~K.}\ \bibnamefont {Kim}},
  \bibinfo {author} {\bibfnamefont {K.~H.}\ \bibnamefont {Ryu}}, \ and\
  \bibinfo {author} {\bibfnamefont {K.}~\bibnamefont {Kang}},\ }\href@noop {}
  {\bibfield  {journal} {\bibinfo  {journal} {Nat. Comm.}\ }\textbf {\bibinfo
  {volume} {7}},\ \bibinfo {pages} {10670} (\bibinfo {year}
  {2016})}\BibitemShut {NoStop}%
\bibitem [{\citenamefont {Kumar}, \citenamefont {Leung},\ and\ \citenamefont
  {Siegel}(2014)}]{vsol_lmno4}%
  \BibitemOpen
  \bibfield  {author} {\bibinfo {author} {\bibfnamefont {N.}~\bibnamefont
  {Kumar}}, \bibinfo {author} {\bibfnamefont {K.}~\bibnamefont {Leung}}, \ and\
  \bibinfo {author} {\bibfnamefont {D.~J.}\ \bibnamefont {Siegel}},\
  }\href@noop {} {\bibfield  {journal} {\bibinfo  {journal} {J. Electrochem.
  Soc.}\ }\textbf {\bibinfo {volume} {161}},\ \bibinfo {pages} {E3059}
  (\bibinfo {year} {2014})}\BibitemShut {NoStop}%
\bibitem [{\citenamefont {Steinmann}\ \emph {et~al.}(2015)\citenamefont
  {Steinmann}, \citenamefont {Michel}, \citenamefont {Schwiedernoch},\ and\
  \citenamefont {Sautet}}]{Steinmann2015a}%
  \BibitemOpen
  \bibfield  {author} {\bibinfo {author} {\bibfnamefont {S.~N.}\ \bibnamefont
  {Steinmann}}, \bibinfo {author} {\bibfnamefont {C.}~\bibnamefont {Michel}},
  \bibinfo {author} {\bibfnamefont {R.}~\bibnamefont {Schwiedernoch}}, \ and\
  \bibinfo {author} {\bibfnamefont {P.}~\bibnamefont {Sautet}},\ }\href
  {\doibase DOI: 10.1039/C5CP00946D} {\bibfield  {journal} {\bibinfo  {journal}
  {Phys. Chem. Chem. Phys.}\ }\textbf {\bibinfo {volume} {17}},\ \bibinfo
  {pages} {13949} (\bibinfo {year} {2015})}\BibitemShut {NoStop}%
\bibitem [{\citenamefont {Sakong}\ \emph {et~al.}(2015)\citenamefont {Sakong},
  \citenamefont {Naderian}, \citenamefont {Mathew}, \citenamefont {Hennig},\
  and\ \citenamefont {Gro{\ss}}}]{vsol_gross}%
  \BibitemOpen
  \bibfield  {author} {\bibinfo {author} {\bibfnamefont {S.}~\bibnamefont
  {Sakong}}, \bibinfo {author} {\bibfnamefont {M.}~\bibnamefont {Naderian}},
  \bibinfo {author} {\bibfnamefont {K.}~\bibnamefont {Mathew}}, \bibinfo
  {author} {\bibfnamefont {R.~G.}\ \bibnamefont {Hennig}}, \ and\ \bibinfo
  {author} {\bibfnamefont {A.}~\bibnamefont {Gro{\ss}}},\ }\href@noop {}
  {\bibfield  {journal} {\bibinfo  {journal} {J. Chem. Phys.}\ }\textbf
  {\bibinfo {volume} {142}},\ \bibinfo {pages} {234107} (\bibinfo {year}
  {2015})}\BibitemShut {NoStop}%
\bibitem [{\citenamefont {Vatti}, \citenamefont {Todorova},\ and\ \citenamefont
  {Neugebauer}(2016)}]{vsol_vatti}%
  \BibitemOpen
  \bibfield  {author} {\bibinfo {author} {\bibfnamefont {A.~K.}\ \bibnamefont
  {Vatti}}, \bibinfo {author} {\bibfnamefont {M.}~\bibnamefont {Todorova}}, \
  and\ \bibinfo {author} {\bibfnamefont {J.}~\bibnamefont {Neugebauer}},\
  }\href@noop {} {\bibfield  {journal} {\bibinfo  {journal} {Langmuir}\
  }\textbf {\bibinfo {volume} {32}},\ \bibinfo {pages} {1027} (\bibinfo {year}
  {2016})}\BibitemShut {NoStop}%
\bibitem [{\citenamefont {Goodpaster}, \citenamefont {Bell},\ and\
  \citenamefont {Head-Gordon}(2016)}]{vsol_goodpaster}%
  \BibitemOpen
  \bibfield  {author} {\bibinfo {author} {\bibfnamefont {J.~D.}\ \bibnamefont
  {Goodpaster}}, \bibinfo {author} {\bibfnamefont {A.~T.}\ \bibnamefont
  {Bell}}, \ and\ \bibinfo {author} {\bibfnamefont {M.}~\bibnamefont
  {Head-Gordon}},\ }\href@noop {} {\bibfield  {journal} {\bibinfo  {journal}
  {J. Phy. Chem. Let.}\ }\textbf {\bibinfo {volume} {7}},\ \bibinfo {pages}
  {1471} (\bibinfo {year} {2016})}\BibitemShut {NoStop}%
\bibitem [{\citenamefont {Sakong}\ and\ \citenamefont
  {Gro{\ss}}(2016)}]{vsol_sakong}%
  \BibitemOpen
  \bibfield  {author} {\bibinfo {author} {\bibfnamefont {S.}~\bibnamefont
  {Sakong}}\ and\ \bibinfo {author} {\bibfnamefont {A.}~\bibnamefont
  {Gro{\ss}}},\ }\href {\doibase 10.1021/acscatal.6b00931} {\bibfield
  {journal} {\bibinfo  {journal} {ACS Cat.}\ }\textbf {\bibinfo {volume} {6}},\
  \bibinfo {pages} {5575} (\bibinfo {year} {2016})}\BibitemShut {NoStop}%
\bibitem [{\citenamefont {Letchworth-Weaver}\ and\ \citenamefont
  {Arias}(2012)}]{ta4}%
  \BibitemOpen
  \bibfield  {author} {\bibinfo {author} {\bibfnamefont {K.}~\bibnamefont
  {Letchworth-Weaver}}\ and\ \bibinfo {author} {\bibfnamefont {T.~A.}\
  \bibnamefont {Arias}},\ }\href {\doibase 10.1103/PhysRevB.86.075140}
  {\bibfield  {journal} {\bibinfo  {journal} {Phys. Rev. B}\ }\textbf {\bibinfo
  {volume} {86}},\ \bibinfo {pages} {075140} (\bibinfo {year}
  {2012})}\BibitemShut {NoStop}%
\bibitem [{\citenamefont {Sundararaman}\ \emph {et~al.}(2012)\citenamefont
  {Sundararaman}, \citenamefont {Gunceler}, \citenamefont {Letchworth-Weaver},\
  and\ \citenamefont {Arias.}}]{ta9}%
  \BibitemOpen
  \bibfield  {author} {\bibinfo {author} {\bibfnamefont {R.}~\bibnamefont
  {Sundararaman}}, \bibinfo {author} {\bibfnamefont {D.}~\bibnamefont
  {Gunceler}}, \bibinfo {author} {\bibfnamefont {K.}~\bibnamefont
  {Letchworth-Weaver}}, \ and\ \bibinfo {author} {\bibfnamefont {T.~A.}\
  \bibnamefont {Arias.}},\ }\href@noop {} {\enquote {\bibinfo {title}
  {{JDFTx}},}\ }\bibinfo {howpublished} {\url{http://jdftx.sourceforge.net}}
  (\bibinfo {year} {2012})\BibitemShut {NoStop}%
\bibitem [{\citenamefont {Gunceler}\ \emph {et~al.}(2013)\citenamefont
  {Gunceler}, \citenamefont {Letchworth-Weaver}, \citenamefont {Sundararaman},
  \citenamefont {Schwarz},\ and\ \citenamefont {Arias}}]{ta11}%
  \BibitemOpen
  \bibfield  {author} {\bibinfo {author} {\bibfnamefont {D.}~\bibnamefont
  {Gunceler}}, \bibinfo {author} {\bibfnamefont {K.}~\bibnamefont
  {Letchworth-Weaver}}, \bibinfo {author} {\bibfnamefont {R.}~\bibnamefont
  {Sundararaman}}, \bibinfo {author} {\bibfnamefont {K.~A.}\ \bibnamefont
  {Schwarz}}, \ and\ \bibinfo {author} {\bibfnamefont {T.~A.}\ \bibnamefont
  {Arias}},\ }\href@noop {} {\bibfield  {journal} {\bibinfo  {journal}
  {Modelling Simul. Mater. Sci. Eng.}\ }\textbf {\bibinfo {volume} {21}},\
  \bibinfo {pages} {074005} (\bibinfo {year} {2013})}\BibitemShut {NoStop}%
\bibitem [{\citenamefont {Sharp}\ and\ \citenamefont {Honig}(1990)}]{honig}%
  \BibitemOpen
  \bibfield  {author} {\bibinfo {author} {\bibfnamefont {K.~A.}\ \bibnamefont
  {Sharp}}\ and\ \bibinfo {author} {\bibfnamefont {B.}~\bibnamefont {Honig}},\
  }\href@noop {} {\bibfield  {journal} {\bibinfo  {journal} {J. Phys. Chem.}\
  }\textbf {\bibinfo {volume} {94}},\ \bibinfo {pages} {7684} (\bibinfo {year}
  {1990})}\BibitemShut {NoStop}%
\bibitem [{\citenamefont {Kohn}\ and\ \citenamefont {Sham}(1965)}]{ks}%
  \BibitemOpen
  \bibfield  {author} {\bibinfo {author} {\bibfnamefont {W.}~\bibnamefont
  {Kohn}}\ and\ \bibinfo {author} {\bibfnamefont {L.~J.}\ \bibnamefont
  {Sham}},\ }\href {\doibase 10.1103/PhysRev.140.A1133} {\bibfield  {journal}
  {\bibinfo  {journal} {Phys. Rev.}\ }\textbf {\bibinfo {volume} {140}},\
  \bibinfo {pages} {A1133} (\bibinfo {year} {1965})}\BibitemShut {NoStop}%
\bibitem [{\citenamefont {Vanderbilt}(1990)}]{uspp}%
  \BibitemOpen
  \bibfield  {author} {\bibinfo {author} {\bibfnamefont {D.}~\bibnamefont
  {Vanderbilt}},\ }\href@noop {} {\bibfield  {journal} {\bibinfo  {journal}
  {Phys. Rev. B}\ }\textbf {\bibinfo {volume} {41}},\ \bibinfo {pages} {7892}
  (\bibinfo {year} {1990})}\BibitemShut {NoStop}%
\bibitem [{\citenamefont {Kresse}\ and\ \citenamefont
  {Joubert}(1999)}]{kresseuspp}%
  \BibitemOpen
  \bibfield  {author} {\bibinfo {author} {\bibfnamefont {G.}~\bibnamefont
  {Kresse}}\ and\ \bibinfo {author} {\bibfnamefont {D.}~\bibnamefont
  {Joubert}},\ }\href {\doibase 10.1103/PhysRevB.59.1758} {\bibfield  {journal}
  {\bibinfo  {journal} {Phys. Rev. B}\ }\textbf {\bibinfo {volume} {59}},\
  \bibinfo {pages} {1758} (\bibinfo {year} {1999})}\BibitemShut {NoStop}%
\bibitem [{\citenamefont {Bl\"ochl}(1994)}]{paw}%
  \BibitemOpen
  \bibfield  {author} {\bibinfo {author} {\bibfnamefont {P.~E.}\ \bibnamefont
  {Bl\"ochl}},\ }\href {\doibase 10.1103/PhysRevB.50.17953} {\bibfield
  {journal} {\bibinfo  {journal} {Phys. Rev. B}\ }\textbf {\bibinfo {volume}
  {50}},\ \bibinfo {pages} {17953} (\bibinfo {year} {1994})}\BibitemShut
  {NoStop}%
\bibitem [{\citenamefont {Santos}\ and\ \citenamefont
  {Schmickler}(2004)}]{Santos2004}%
  \BibitemOpen
  \bibfield  {author} {\bibinfo {author} {\bibfnamefont {E.}~\bibnamefont
  {Santos}}\ and\ \bibinfo {author} {\bibfnamefont {W.}~\bibnamefont
  {Schmickler}},\ }\href {\doibase
  https://doi.org/10.1016/j.cplett.2004.10.072} {\bibfield  {journal} {\bibinfo
   {journal} {Chem. Phys. Lett.}\ }\textbf {\bibinfo {volume} {400}},\ \bibinfo
  {pages} {26 } (\bibinfo {year} {2004})}\BibitemShut {NoStop}%
\bibitem [{\citenamefont {Trasatti}\ and\ \citenamefont
  {Lust}(1999)}]{Trasatti1999}%
  \BibitemOpen
  \bibfield  {author} {\bibinfo {author} {\bibfnamefont {S.}~\bibnamefont
  {Trasatti}}\ and\ \bibinfo {author} {\bibfnamefont {E.}~\bibnamefont
  {Lust}},\ }in\ \href {\doibase 10.1007/0-306-46917-0_1} {\emph {\bibinfo
  {booktitle} {Modern Aspects of Electrochemistry}}},\ \bibinfo {editor}
  {edited by\ \bibinfo {editor} {\bibfnamefont {R.~E.}\ \bibnamefont {White}},
  \bibinfo {editor} {\bibfnamefont {J.~O.}\ \bibnamefont {Bockris}}, \ and\
  \bibinfo {editor} {\bibfnamefont {B.~E.}\ \bibnamefont {Conway}}}\ (\bibinfo
  {publisher} {Springer US},\ \bibinfo {address} {Boston, MA},\ \bibinfo {year}
  {1999})\ pp.\ \bibinfo {pages} {1--215}\BibitemShut {NoStop}%
\bibitem [{\citenamefont {Perdew}, \citenamefont {Burke},\ and\ \citenamefont
  {Ernzerhof}(1996)}]{pbe}%
  \BibitemOpen
  \bibfield  {author} {\bibinfo {author} {\bibfnamefont {J.~P.}\ \bibnamefont
  {Perdew}}, \bibinfo {author} {\bibfnamefont {K.}~\bibnamefont {Burke}}, \
  and\ \bibinfo {author} {\bibfnamefont {M.}~\bibnamefont {Ernzerhof}},\
  }\href@noop {} {\bibfield  {journal} {\bibinfo  {journal} {Phys. Rev. Lett.}\
  }\textbf {\bibinfo {volume} {77}},\ \bibinfo {pages} {3865} (\bibinfo {year}
  {1996})}\BibitemShut {NoStop}%
\bibitem [{\citenamefont {Pajkossy}\ and\ \citenamefont
  {Kolb}(2001)}]{Pajkossy2001}%
  \BibitemOpen
  \bibfield  {author} {\bibinfo {author} {\bibfnamefont {T.}~\bibnamefont
  {Pajkossy}}\ and\ \bibinfo {author} {\bibfnamefont {D.}~\bibnamefont
  {Kolb}},\ }\href {\doibase http://dx.doi.org/10.1016/S0013-4686(01)00597-7}
  {\bibfield  {journal} {\bibinfo  {journal} {Electrochim. Acta}\ }\textbf
  {\bibinfo {volume} {46}},\ \bibinfo {pages} {3063 } (\bibinfo {year}
  {2001})}\BibitemShut {NoStop}%
\bibitem [{\citenamefont {Steinmann}\ and\ \citenamefont
  {Sautet}(2016)}]{Steinmann2016a}%
  \BibitemOpen
  \bibfield  {author} {\bibinfo {author} {\bibfnamefont {S.~N.}\ \bibnamefont
  {Steinmann}}\ and\ \bibinfo {author} {\bibfnamefont {P.}~\bibnamefont
  {Sautet}},\ }\href {\doibase 10.1021/acs.jpcc.6b01938} {\bibfield  {journal}
  {\bibinfo  {journal} {J Phys Chem C}\ }\textbf {\bibinfo {volume} {120}},\
  \bibinfo {pages} {5619} (\bibinfo {year} {2016})}\BibitemShut {NoStop}%
\bibitem [{\citenamefont {Tran}\ \emph {et~al.}(2016)\citenamefont {Tran},
  \citenamefont {Xu}, \citenamefont {Radhakrishnan}, \citenamefont {Winston},
  \citenamefont {Sun}, \citenamefont {Persson},\ and\ \citenamefont
  {Ong}}]{MP_wulff}%
  \BibitemOpen
  \bibfield  {author} {\bibinfo {author} {\bibfnamefont {R.}~\bibnamefont
  {Tran}}, \bibinfo {author} {\bibfnamefont {Z.}~\bibnamefont {Xu}}, \bibinfo
  {author} {\bibfnamefont {B.}~\bibnamefont {Radhakrishnan}}, \bibinfo {author}
  {\bibfnamefont {D.}~\bibnamefont {Winston}}, \bibinfo {author} {\bibfnamefont
  {W.}~\bibnamefont {Sun}}, \bibinfo {author} {\bibfnamefont {K.~A.}\
  \bibnamefont {Persson}}, \ and\ \bibinfo {author} {\bibfnamefont {S.~P.}\
  \bibnamefont {Ong}},\ }\href@noop {} {\bibfield  {journal} {\bibinfo
  {journal} {Sci. data}\ }\textbf {\bibinfo {volume} {3}},\ \bibinfo {pages}
  {160080} (\bibinfo {year} {2016})}\BibitemShut {NoStop}%
\bibitem [{\citenamefont {Mathew}\ \emph {et~al.}(2016)\citenamefont {Mathew},
  \citenamefont {Singh}, \citenamefont {Gabriel}, \citenamefont {Choudhary},
  \citenamefont {Sinnott}, \citenamefont {Davydov}, \citenamefont {Tavazza},\
  and\ \citenamefont {Hennig}}]{Mathew2016mpinterfaces}%
  \BibitemOpen
  \bibfield  {author} {\bibinfo {author} {\bibfnamefont {K.}~\bibnamefont
  {Mathew}}, \bibinfo {author} {\bibfnamefont {A.~K.}\ \bibnamefont {Singh}},
  \bibinfo {author} {\bibfnamefont {J.~J.}\ \bibnamefont {Gabriel}}, \bibinfo
  {author} {\bibfnamefont {K.}~\bibnamefont {Choudhary}}, \bibinfo {author}
  {\bibfnamefont {S.~B.}\ \bibnamefont {Sinnott}}, \bibinfo {author}
  {\bibfnamefont {A.~V.}\ \bibnamefont {Davydov}}, \bibinfo {author}
  {\bibfnamefont {F.}~\bibnamefont {Tavazza}}, \ and\ \bibinfo {author}
  {\bibfnamefont {R.~G.}\ \bibnamefont {Hennig}},\ }\href {\doibase
  10.1016/j.commatsci.2016.05.020} {\bibfield  {journal} {\bibinfo  {journal}
  {Comp. Mater. Sci.}\ }\textbf {\bibinfo {volume} {122}},\ \bibinfo {pages}
  {183} (\bibinfo {year} {2016})}\BibitemShut {NoStop}%
\end{thebibliography}%

\end{document}